\documentclass[aps,pra,10pt,twocolumn,superscriptaddress,floatfix]{revtex4-1}
\usepackage{dcolumn}
\usepackage{amsmath}
\usepackage{amssymb}
\usepackage{bm}
\usepackage{times}
\usepackage{graphicx}
\usepackage{xcolor}

\usepackage{braket}
\usepackage{tikz}

\newcolumntype{.}{D{x}{}{-1}}

\newcommand*{\centt}[1]{\multicolumn{1}{c}{#1}}
\newcolumntype{w}[1]{D{.}{.}{#1}}

\definecolor{corcolor}{rgb}{0.2,0.2,1}

\allowdisplaybreaks
\begin{document}
\preprint{Version 1.1}

\title{Nonadiabatic relativistic correction in  H$_2$, D$_2$, and HD}
\author{Pawe\l\ Czachorowski}
\affiliation{Faculty of Physics, University of Warsaw, Pasteura 5, 02-093 Warsaw, Poland}

\author{Mariusz Puchalski} 
\affiliation{Faculty of Chemistry, Adam Mickiewicz University,
             Umultowska 89b, 61-614 Pozna{\'n}, Poland}

\author{Jacek Komasa} 
\affiliation{Faculty of Chemistry, Adam Mickiewicz University,
             Umultowska 89b, 61-614 Pozna{\'n}, Poland}

\author{Krzysztof Pachucki}
\affiliation{Faculty of Physics, University of Warsaw, Pasteura 5, 02-093 Warsaw, Poland}

\date{\today}

\begin{abstract} 
  \noindent We calculate the nonadiabatic relativistic correction to rovibrational energy levels of H$_2$, D$_2$,
  and HD molecules using the nonadiabatic perturbation theory. This approach allows one to obtain nonadiabatic corrections
  to all the molecular levels with the help of a single effective potential. The obtained results are in very good agreement
  with the previous direct calculation of nonadiabatic relativistic effects for dissociation energies
  and resolve the reported discrepancies of theoretical predictions with recent experimental results. 
\end{abstract} 
\maketitle 
 
\section{Introduction}
The hydrogen molecule has not yet been used for deter-mination of fundamental physical constants,
unlike atomic hydrogen. This is due to difficulties in
accurate solution of the molecular Schr\"odinger equation and an inherence of an electron correlation, 
combined with relativistic, quantum electrodynamic, and nonadiabatic effects.
At the precision level of $10^{-7}$ cm$^{-1}$ vibrational excitations are sensitive to
uncertainties in the electron-proton mass ratio, in the nuclear charge radii, and in the Rydberg constant.
Therefore, from sufficiently accurate theoretical predictions and corresponding measurements
one can obtain those fundamental physical constants.

To deal with the problem of accurate calculation of molecular levels in a systematic manner,
one employs the nonrelativistic quantum electrodynamic (NRQED) approach \cite{NRQED},
which is a perturbation theory that can be made to agree with the full quantum electrodynamics (QED)
up to an arbitrary order in the fine structure constant $\alpha$. 
It assumes an expansion of the binding energy in $\alpha$ 
\begin{equation}
E(\alpha) = \alpha^2E^{(2)} + \alpha^4E^{(4)} + \alpha^5E^{(5)} + \alpha^6E^{(6)} + \alpha^7E^{(7)} + \mathcal{O}(\alpha^8), 
\end{equation}
where $E^{(n)}$ is a contribution of order $\alpha^n\,m$ and may include powers of $\ln\alpha$.
Each $E^{(n)}$ can be expressed as an expectation
value of some effective Hamiltonian with the nonrelativistic wave function.
These expansion terms can, in turn, be expanded further -- in another series of the electron-nuclear mass ratio --
to obtain the contributions of the Born-Oppenheimer, adiabatic, and nonadiabatic effects.
These contributions can be calculated within the so-called nonadiabatic perturbation theory (NAPT) \cite{NAPT}.

Significant progress has been achieved in recent years
by the accurate ($\sim 10^{-7}$ cm$^{-1}$) direct solution of the four-body Schr\"odinger equation \cite{komasa_2018},
while the calculations of relativistic $(\alpha^4m)$, quantum electrodynamic $(\alpha^5m)$,
and higher order quantum electrodynamic $(\alpha^6m)$ corrections were performed within the Born-Oppenheimer
(BO) approximation. The resulting theoretical predictions
happened to be in about $3\,\sigma$ disagreement with recent experimental results.
It was suggested, for the resolution of these discrepancies,
that an estimate of relativistic nonadiabatic corrections by the factor of
the electron-nucleus mass ratio might not be correct.
Indeed, very recent fully nonadiabatic calculations for the ground molecular state \cite{Yan, MariuszNa, Yan2}
have demonstrated that these corrections are about 10 times larger than expected and explain
the apparent discrepancy with measured dissociation energies for H$_2$ and D$_2$. For HD, however, a
$2\,\sigma$ discrepancy remains and this requires further investigations.

In this paper we provide the results for the relativistic nonadiabatic correction
obtained with a perturbative approach based on NAPT. More importantly,
this method retains the key benefit of the adiabatic approximation --
the existence of the potential energy curve, which, calculated once for a given electronic state,
can be utilized to easily obtain all rovibrational energies. The obtained results are
in a very good agreement with the direct calculation of nonadiabatic relativistic correction for
the ground molecular state and explain almost all previously reported discrepancies for various transition energies.

\section{Derivation of Formulas}
We pass now to the derivation of formulas for the nonadiabatic relativistic correction.
The Schr\"odinger equation for a bielectronic, binuclear molecule, written in a center-of-mass frame,
with the origin in the geometric center of the nuclei, is
\begin{align}
(H+H_\mathrm{n}-E^{(2)})\ket{\Psi(\vec{r}_1,\vec{r}_2,\vec{R})}&=0, \label{fullschr}
\end{align}
where
\begin{align}
H&=-\frac{1}{2}\left(\vec{\nabla}^2_1+\vec{\nabla}^2_2\right)+V,\\
V&=-\frac{1}{r_{1A}}-\frac{1}{r_{1B}}-\frac{1}{r_{2A}}-\frac{1}{r_{2B}}+\frac{1}{r_{12}}+\frac{1}{R},\\
H_\mathrm{n}&=-\frac{1}{2\mu_\mathrm{n}}\left(\vec{\nabla}_R^2+\vec{\nabla}_\mathrm{el}^2\right)+
             \left(\frac{1}{M_A}-\frac{1}{M_\mathrm{B}}\right)\vec{\nabla}_R\vec{\nabla}_\mathrm{el}, \label{hen}
\end{align}
and where  $1$, $2$ indices denote electrons, $A$, $B$ indicate nuclei, the nuclear reduced mass
$\mu_\mathrm{n}=M_AM_\mathrm{B}/(M_A+M_\mathrm{B})$,
$\vec{R} = \vec{R}_A-\vec{R}_B$, and $\vec{\nabla}_\mathrm{el}=(\vec{\nabla}_1+\vec{\nabla}_2)/2$.
In homonuclear molecules, such as H$_2$ or D$_2$, the last term in $H_\mathrm{n}$ vanishes, whereas in HD it is present.
However, it is neglected anyway because it contributes in the second order of the electron-nuclear mass ratio,
while our calculations of relativistic nonadiabatic corrections are performed only to the first order.
The magnitude of these neglected terms is estimated in Sec. VII and verified against nonperturbative calculations
for the ground molecular state \cite{Yan, MariuszNa, Yan2}.

The function $\Psi(\vec{r}_1,\vec{r}_2,\vec{R})$ in Eq. (\ref{fullschr})
is the solution of the full Schr\"{o}dinger equation for H$_2$, describing both the electrons and the nuclei.
Here, however, we employ the NAPT formalism and represent the wave function as
\begin{align}
  \Psi(\vec{r}_1,\vec{r}_2,\vec{R})&=\psi(\vec{r}_1,\vec{r}_2)\chi(\vec{R})+\delta\Psi_{na}(\vec{r}_1,\vec{r}_2,\vec{R}),
  \label{naptwave}
\end{align}
where the matrix element in the electron space vanishes $\braket{\delta\Psi_{na}|\psi}=0$, and
$\psi(\vec{r}_1,\vec{r}_2)$ is an eigenfunction of the electronic Schr\"odinger equation
\begin{align}
H\ket{\psi}&=\mathcal{E}^{(2,0)}(R)\ket{\psi},\label{elecschro}
\end{align}
with the eigenvalue dependent on the internuclear distance $R$.
The function $\chi$ satisfies the following nuclear equation
\begin{align}
\Bigl[-\frac{\nabla^2_R}{2\,\mu_{\rm n}} + {\cal E}^{(2,0)}(R) - E^{(2,0)}\Bigr]\,\chi(\vec R) = 0.
\end{align}
For convenience, from now on we will denote ${\mathcal{E}(R)\equiv\mathcal{E}^{(2,0)}(R)}$.
The leading finite nuclear correction is given by
\begin{align}
E^{(2,1)}=\langle\chi|{\cal E}^{(2,1)}(R)|\chi\rangle,
\end{align}
where $\mathcal{E}^{(2,1)}$ is the expectation value of $H_\mathrm{n}$, known as the adiabatic correction,
\begin{align}
\mathcal{E}^{(2,1)}(R)&=\braket{\psi|H_\mathrm{n}|\psi} \nonumber\\
&=\frac{1}{2\mu_\mathrm{n}}\braket{\vec{\nabla}_R\psi|\vec{\nabla}_R\psi}+\frac{1}{2\mu_\mathrm{n}}\braket{\vec{\nabla}_\mathrm{el}\psi|\vec{\nabla}_\mathrm{el}\psi}. \label{adiab}
\end{align}
This correction is known with a high accuracy from Ref. \cite{adiab}.
The remainder $\delta\Psi_{\rm na}$ will not be needed because we calculate here only the leading corrections
in the electron nuclear mass ratio.

In analogy to the nonrelativistic energies, the relativistic BO correction is an expectation value of the Breit-Pauli Hamiltonian
with the electronic wave function
\begin{align}
  \mathcal{E}^{(4,0)}(R)&=\braket{\psi|H^{(4,0)}|\psi},
\end{align}
where
\begin{align}
H^{(4,0)}&=-\frac{p^4_1+p^4_2}{8} -\frac{1}{2}p_1^i\left(\frac{\delta^{ij}}{r_{12}}+\frac{r_{12}^ir_{12}^j}{r^3_{12}}\right)p_2^j +\pi \delta^3(r_{12}) \nonumber \\
&+\frac{\pi}{2}\left(\delta^3(r_{1A})+\delta^3(r_{2A})+\delta^3(r_{1B})+\delta^3(r_{2B})\right).\label{hbp}
\end{align}
It has recently been recalculated with a high accuracy in Ref. \cite{Mariuszrel}.
The topic of this work is a combined, nonadiabatic--relativistic correction $\mathcal{E}^{(4,1)}(R)$, which is
represented as a sum of three terms
\begin{align}
  \mathcal{E}^{(4,1)}(R)&=\frac{1}{\mu_\mathrm{n}}\left[\mathcal{E}_1^{(4,1)}(R) +
    \mathcal{E}_2^{(4,1)}(R) + \mathcal{E}_3^{(4,1)}(R)\right],\label{reladcomplete}
\end{align}
where
\begin{align}
\mathcal{E}_1^{(4,1)}(R)&=\braket{\vec{\nabla}_R\psi_\mathrm{rel}|\vec{\nabla}_R\psi}, \label{corra1} \\
\mathcal{E}_2^{(4,1)}(R)&=-\braket{\psi_\mathrm{rel}|\vec{\nabla}^2_\mathrm{el}|\psi}, \label{corra2}\\
\mathcal{E}_3^{(4,1)}(R)&=\mu_\mathrm{n}\braket{\psi|H_M^{(4,1)}|\psi},\label{corra3}
\end{align}
and where
\begin{align}
\ket{\psi_\mathrm{rel}}&=\frac{1}{(\mathcal{E}-H)'}H^{(4,0)}\ket{\psi},\\
H_M^{(4,1)}&=-\sum_{a=1,2}\sum_{N=A,B}\frac{1}{2\,M_N}\nabla_a^{i}\left(\frac{\delta^{ij}}{r_{aN}}+\frac{r_{aN}^ir_{aN}^j}{r^3_{aN}}\right)\nabla_N^{j}.
\end{align}
In our coordinate system, with the center-of-mass at rest and the origin in the geometric center of the nuclei,
$H_{M}^{(4,1)}$ takes the form
\begin{align}
  &H_{M}^{(4,1)}=\label{hm} \\ &
              -\frac{1}{4\mu_\mathrm{n}}\sum_{a=1,2}\nabla_a^{i}\left(\frac{\delta^{ij}}{r_{aA}}
               +\frac{r_{aA}^ir_{aA}^j}{r^3_{aA}}-\frac{\delta^{ij}}{r_{aB}}-\frac{r_{aB}^ir_{aB}^j}{r^3_{aB}}\right)\nabla_R^{j}\nonumber\\
              &+\frac{1}{4\mu_\mathrm{n}}\sum_{a=1,2}\nabla_a^{i}\left(\frac{\delta^{ij}}{r_{aA}}
               +\frac{r_{aA}^ir_{aA}^j}{r^3_{aA}}+\frac{\delta^{ij}}{r_{aB}}+\frac{r_{aB}^ir_{aB}^j}{r^3_{aB}}\right)\nabla_\mathrm{el}^{j}. \nonumber
\end{align}
Potentials $\mathcal{E}_1^{(4,1)}$ and $\mathcal{E}_3^{(4,1)}$ involve $\vec{\nabla}_R$ -- a gradient with respect to the internuclear vector $\vec R$.
It should be handled properly, which is described in the next section.

\section{Nuclear gradients}
We consider at first $\vec{\nabla}_R$ acting on the nonrelativistic wave function.
It can be obtained by differentiation of the Shr\"odinger equation
\begin{align}
  \vec{\nabla}_R\ket{\psi}&=\frac{1}{(\mathcal{E}-H)'}\vec{\nabla}_R(V)\ket{\psi} = \vec n\,\ket{\psi_\mathrm{a}} - \vec n\times\ket{\vec{\psi}_\mathrm{a}}, \label{proj1}
\end{align}
where $\vec{n}={\vec{R}}/{R}$ and
\begin{align}
\ket{\psi_\mathrm{a}}&= \vec{n}\cdot\vec{\nabla}_R\ket{\psi}=\frac{\vec{n}}{(\mathcal{E}-H)'}\vec{\nabla}_R(V)\ket{\psi}, \label{psia}\\
\ket{\vec{\psi}_\mathrm{a}}&= \vec{n}\times\vec{\nabla}_R\ket{\psi}=\frac{1}{(\mathcal{E}-H)'}\vec{n}\times\vec{\nabla}_R(V)\ket{\psi}, \label{psipa}
\end{align}
and where
\begin{align}
\vec{\nabla}_R(V)&=\frac{1}{2}\left(-\frac{\vec{r}_{1A}}{r_{1A}^3}+\frac{\vec{r}_{1B}}{r_{1B}^3}-\frac{\vec{r}_{2A}}{r_{2A}^3}+\frac{\vec{r}_{2B}}{r_{2B}^3}\right)-\frac{\vec{R}}{R^3}\,.
\end{align}
The analogous derivative of $\psi_{\rm rel}$ would be quite complicated, and therefore 
we recast the expression for $\mathcal{E}_1^{(4,1)}$ into a more tractable form as follows
\begin{align}
\mathcal{E}_1^{(4,1)}(R)&=\vec{\nabla}_R\braket{\psi_\mathrm{rel}|\vec{\nabla}_R\psi}-\braket{\psi_\mathrm{rel}|\vec{\nabla}_R^2|\psi}.\label{ea1}
\end{align}
The first term above can be evaluated by numerical differentiation
\begin{align}
  \vec{\nabla}_R\braket{\psi_\mathrm{rel}|\vec{\nabla}_R|\psi}&=
  \frac{1}{R^2}\frac{\partial}{\partial R}\left(R^2\braket{\psi_\mathrm{rel}|\psi_\mathrm{a}}\right)\,.
\end{align}
In practical application, it is done by polynomial interpolation of $\braket{\psi_\mathrm{rel}|\psi_\mathrm{a}}$
and a subsequent derivative.

The second term in Eq. (\ref{ea1}) is obtained as follows
\begin{align}
\vec{\nabla}_R^2\ket{\psi}&=\ket{\psi_\mathrm{na}} + c\ket{\psi}, \label{psina}
\end{align}
where
\begin{align}
\ket{\psi_\mathrm{na}}=&\frac{1}{(\mathcal{E}-H)'}\left[\vec{\nabla}^2_R(V)\ket{\psi}\right. \nonumber\\
&\left.+ 2\,\vec{\nabla}_R(V-\mathcal{E})\frac{1}{(\mathcal{E}-H)'}\vec{\nabla}_R(V)\ket{\psi}\right]
\end{align}
is orthogonal to $\ket{\psi}$, and
\begin{align}
\vec{\nabla}^2_R(V)=&\ \pi\left[\delta^3(r_{1A})+\delta^3(r_{2A})+\delta^3(r_{1B})+\delta^3(r_{2B})\right], \\
  \vec{\nabla}_R(\mathcal{E})=&\ \braket{\psi|\vec{\nabla}_R(V)|\psi}.
\end{align}
The term  $c\,\ket{\psi}$ in Eq.  (\ref{psina})  would appear next to a reduced resolvent
from $\ket{\psi_\mathrm{rel}}$ in Eq. (\ref{ea1}), so it does not contribute.
Next, we decompose the second resolvent into $\Sigma$ and $\Pi$ parts.
Such partition enables one to represent the resolvent in states of specific symmetry,
which simplifies the numerical implementation. Gathering it all together,
we obtain the following transformed form of Eq. (\ref{ea1})
\begin{align}
  \mathcal{E}_1^{(4,1)}(R)&=
  \frac{1}{R^2}\frac{\partial}{\partial R}\left(R^2\braket{\psi_\mathrm{rel}|\psi_\mathrm{a}}\right)-\braket{\psi_\mathrm{rel}|\psi_\mathrm{na}},
\end{align}
where
\begin{align}
 \ket{\psi_\mathrm{na}}=&\frac{1_{\Sigma^+}}{(\mathcal{E}-H)'}\,\vec{\nabla}^2_R(V)\ket{\psi} \nonumber \\
      &+ 2\,\frac{1_{\Sigma^+}}{(\mathcal{E}-H)'}\vec{n}\cdot\vec{\nabla}_R(V-{\cal E})\ket{\psi_\mathrm{a}}\ \nonumber \\
      &+ 2\,\frac{1_{\Sigma^+}}{(\mathcal{E}-H)'}\vec{n}\times\vec{\nabla}_R(V)\ket{\vec{\psi}_\mathrm{a}}.
\end{align}
The analogous separation of intermediate states of definite symmetry is performed for Eq. (\ref{hm})
\begin{align}
&\ \mathcal{E}_3^{(4,1)}(R) = \frac{1}{4}\sum_{a=1,2}\biggl[ \\
    &\ \braket{\psi|\nabla^i_{a}\left(\frac{\delta^{ij}}{r_{aA}}
      +\frac{r_{aA}^ir_{aA}^j}{r^3_{aA}}+\frac{\delta^{ij}}{r_{aB}}+\frac{r_{aB}^ir_{aB}^j}{r^3_{aB}}\right)\nabla^j_\mathrm{el}|\psi} \nonumber \\
    &-\braket{\psi|n^j\nabla_a^{i}\left(\frac{\delta^{ij}}{r_{aA}}
      +\frac{r_{aA}^ir_{aA}^j}{r^3_{aA}}-\frac{\delta^{ij}}{r_{aB}}-\frac{r_{aB}^ir_{aB}^j}{r^3_{aB}}\right)|\psi_\mathrm{a}} \nonumber\\
    &-\braket{\psi|\epsilon^{mkj}n^k\nabla_a^{i}\left(\frac{\delta^{ij}}{r_{aA}}
      +\frac{r_{aA}^ir_{aA}^j}{r^3_{aA}}-\frac{\delta^{ij}}{r_{aB}}-\frac{r_{aB}^ir_{aB}^j}{r^3_{aB}}\right)|\psi^m_\mathrm{a}}\biggr], \nonumber
\end{align}
while $\mathcal{E}_2^{(4,1)}$ does not need any further transformations.

\section{Regularization}
The Breit-Pauli Hamiltonian contains singular type operators, like Dirac $\delta$ and $p^4$,
whose matrix elements have slow numerical convergence. For this reason we perform a regularization
that is based on various expectation value identities \cite{Drachman} (see also \cite{Cencek}).
In the case of Gaussian basis, due to a poor representation of the wave function at coalescence points,
the regularization improves the convergence dramatically \cite{Mariuszrel}. These identities are the following
\begin{align}
  4\pi\delta^3(r_{1A})=&\ \frac{4}{r_{1A}}(\mathcal{E}-V)-\vec p_1\frac{2}{r_{1A}}\,\vec p_1 -
  \vec p_2\frac{2}{r_{1A}}\,\vec p_2
 \nonumber \\ &\ +\left\{\frac{2}{r_{1A}},H-\mathcal{E}\right\},\\
 p_1^4+p_2^4= &\ 4(\mathcal{E}-V)^2-2p_1^2p_2^2+4(H-\mathcal{E})^2
 \nonumber \\ &\ +4\left\{\mathcal{E}-V,H-\mathcal{E}\right\}.
\end{align}
Furthermore, should the $p_1^2p_2^2$ in the above expression act on a wave function that satisfies
the Kato's cusp condition, the arising $\delta^3(r_{12})$ function cancels out exactly with that 
from the Breit-Pauli Hamiltonian and the remainder will be denoted by $\tilde p_1^2\,\tilde p_2^2$.
This regularization has been already employed in calculations of the BO relativistic
corrections \cite{Mariuszrel}. The only, but important, difference is that now the Breit-Pauli Hamiltonian
acts on a wave function other than the reference state's, and subsequently the terms in the anticommutators
cannot be neglected.

After making use of the above formulas, we obtain
\begin{align}
H^{(4,0)}=&\left[H^{(4,0)}\right]_r+\left\{Q,H-\mathcal{E}\right\} - \frac{1}{2}\,(H-\mathcal{E})^2,\label{wave1}\\
\left[H^{(4,0)}\right]_r=&-\frac{1}{2}(\mathcal{E}-V)\left(\mathcal{E}-\frac{1}{R}-\frac{1}{r_{12}}\right)\nonumber\\
&+\frac{1}{4}(\tilde{p}_1^2\tilde{p}_2^2+p_1\tilde{V}p_1+p_2\tilde{V}p_2) \nonumber \\
&-\frac{1}{2}p_1^i\left(\frac{\delta^{ij}}{r_{12}}+\frac{r_{12}^ir_{12}^j}{r_{12}^3}\right)p_2^j 
\end{align}
and
\begin{align}
\vec{\nabla}^2_R(V)&=\left[\vec{\nabla}^2_R(V)\right]_r-\frac{1}{2}\left\{\tilde{V},H-\mathcal{E}\right\},\label{nabla2}\\
\left[\vec{\nabla}^2_R(V)\right]_r&=-(\mathcal{E}-V)\tilde{V}+\frac{1}{2}\,p_1\tilde{V}p_1+\frac{1}{2}\,p_2\tilde{V}p_2,
\end{align}
where
\begin{align}
Q=&-\frac{1}{2}(\mathcal{E}-V)-\frac{1}{4}\,\tilde V,\\
\tilde{V}=&-\frac{1}{r_{1A}}-\frac{1}{r_{2A}}-\frac{1}{r_{1B}}-\frac{1}{r_{2B}}.
\end{align}

\section{Numerical calculations}
The calculations were performed with a variational wave function represented as a linear combination
\begin{align}
\ket{\psi}&=\sum_ic_i\psi_i(\vec{r}_1,\vec{r}_2),\\
\psi_i&=(1+\hat{\imath})(1+P_{1\leftrightarrow 2})\phi_i(\vec{r}_1,\vec{r}_2),
\end{align}
where $\hat{\imath}$ is an inversion operator and $P_{1\leftrightarrow 2}$ exchanges the electrons.
The basis functions $\phi_i(\vec{r}_1,\vec{r}_2)$ are of the explicitly correlated Gaussian (ECG) type
\begin{align}
\phi_{\Sigma^+}&=e^{-a_{1A}r_{1A}^2-a_{1B}r_{1B}^2-a_{2A}r_{2A}^2-a_{2B}r_{2B}^2-a_{12}r_{12}^2},\label{ECG}\\
\vec{\phi}_{\Pi}&= \vec{n}\times\vec{r}_{1A}\,\phi_{\Sigma^+},
\end{align}
where the parameters $a_{1A}$, $a_{2A}$, $a_{1B}$, $a_{2B}$ and $a_{12}$ were optimized individually for each basis function. In addition, for the ground-state wave function we employed the so-called rECG basis
\begin{align}
\phi'_{\Sigma^+}&=(1+r_{12}/2)\,e^{-a_{1A}r_{1A}^2-a_{1B}r_{1B}^2-a_{2A}r_{2A}^2-a_{2B}r_{2B}^2-a_{12}r_{12}^2}, \label{rECG}
\end{align}
which satisfies exactly the inter-electronic cusp condition. It significantly improves the numerical convergence
and allows for algebraic cancellation of the $\delta^3(r_{12})$ term from the Breit-Pauli Hamiltonian.
Both kinds of Gaussian bases have been used before in Refs. \cite{MariuszNa, Mariuszrel}.
All the integrals can be performed either analytically or, in the worst case,
numerically with fast extended Gaussian quadrature \cite{gaussext}.
As a consequence, we can achieve high accuracy with a reasonable low computational cost.

The method for performing integrals has been already described extensively in \cite{ma6,Mariuszrel},
but -- for the sake of completeness -- we repeat the main formulas.
All the matrix elements needed can be written down as linear combinations of $f$'s
\begin{eqnarray}
f(n_1,n_2,n_3,n_4,n_5) &=& \frac{1}{\pi^3}\int d^3 r_1 \int d^3 r_2 \, 
r_{1A}^{n_1} r_{1B}^{n_2} r_{2A}^{n_3} r_{2B}^{n_4} r_{12}^{n_5}
\nonumber \\ && \hspace*{-10ex}
\times e^{-c_{1A}\,r_{1A}^2 -c_{1B}\,r_{1B}^2 -c_{2A}\,r_{2A}^2 -c_{2B}\,r_{2B}^2 - c_{12}\,r_{12}^2 }. 
\label{27}
\end{eqnarray}
The integrals with even powers of the inter-particle distance can be obtained by differentiation over the variational parameters of the 'master' integral
\begin{eqnarray}
f(0,0,0,0,0) = A^{-3/2} e^{-R^2 \frac{B}{A}}, \label{29}
\label{basis_int}
\end{eqnarray}
where
\begin{eqnarray}
A &=& (c_{1A} + c_{1B} + c_{12}) (c_{2A} + c_{2B} + c_{12}) - c_{12}^2, \label{30}\\
B &=&   (c_{1B} + c_{1A}) c_{2A} c_{2B} + c_{1A} c_{1B} (c_{2A} + c_{2B}) 
\nonumber \\ &&
+ c_{12} (c_{1A} + c_{2A}) (c_{1B} + c_{2B}). \label{31}
\end{eqnarray}
If one of the $n_k$ indices is odd, the ECG integrals can also be obtained analytically by differentiation
of other master integrals. As an example, the master integral for $n_1 = -1$ is
\begin{equation}
f(-1,0,0,0,0) = \frac{1}{A \sqrt{A_1}}\,e^{-R^2 \frac{B}{A}}
F \bigg[R^2\bigg( \frac{B_1}{A_1} - \frac{B}{A}\bigg) \bigg], \label{32}
\end{equation}
where $A_1 = \partial_{c_{1A}} \, A$, $B_1 = \partial_{c_{1A}} \,B$, and $F(x) = {\rm erf}(x)/x$.
Molecular ECG integrals, as opposed to the atomic case, have no known analytic
form when two or more $n_k$ indices are odd. In this case, we use the quadrature adapted
to the end-point logarithmic singularity \cite{gaussext}, which has fast numerical convergence.

Eventually, for a given basis size $N$, we had to optimize eight different sets.
The first two, with and without the cusp, are for the ground electronic state and correspond to optimization of the ground state energy ${\cal E}(R) = \langle\psi|H|\psi\rangle$.
The next four basis sets are for intermediate states with the following matrix elements
\begin{align}
  & \bra{\psi}[H^{(4,0)}]_r\frac{1}{(\mathcal{E}-H)'}[H^{(4,0)}]_r\ket{\psi}, \\
  & \bra{\psi}\vec{n}\cdot\vec{\nabla}_R(V-\mathcal{E})\frac{1}{(\mathcal{E}-H)'}
              \vec{n}\cdot\vec{\nabla}_R(V-\mathcal{E}) \ket{\psi}, \\
  & \bra{\psi}\vec{n}\times\vec{\nabla}_R(V)\frac{1}{(\mathcal{E}-H)}\vec{n}\times\vec{\nabla}_R(V)\ket{\psi}, \\
  & \bra{\psi}[\vec{\nabla}^2_R(V)]_r\frac{1}{(\mathcal{E}-H)'}[\vec{\nabla}^2_R(V)]_r\ket{\psi},  
\end{align}  
which can be directly optimized. The last two basis sets are for intermediate states with
\begin{align}
  & \bra{\psi_{\rm a}}\vec{n}\cdot\vec{\nabla}_R(V-\mathcal{E})\frac{1}{(\mathcal{E}-H)'}
              \vec{n}\cdot\vec{\nabla}_R(V-\mathcal{E}) \ket{\psi_{\rm a}}, \\
  & \bra{\vec\psi_\mathrm{a}}\vec{n}\times\vec{\nabla}_R(V)\frac{1}{(\mathcal{E}-H)'}
                                      \vec{n}\times\vec{\nabla}_R(V)\ket{\vec\psi_\mathrm{a}}.
\end{align}  
To ensure proper subtraction of the ground state from the reduced resolvents,
we extended each $\Sigma_+$ basis with a fixed sector consisting of $N/2$ basis functions
optimized for the ground state without a cusp.
Its nonlinear variational parameters were kept constant and were not further optimized.

The calculations were performed for three different basis sizes: $N=128$, $256$, $512$,
to observe numerical convergence and estimate the corresponding uncertainty.
The electronic $\mathcal{E}^{(4,1)}(R)$ potential was calculated for $59$ points in the range of $0$--$8$ a.u.
Results are presented in Table \ref{Rs} and plotted in Fig. 1. The exact value at $R=0$, {$2\mu_\mathrm{n}\mathcal{E}^{(4,1)}(0)=-1.079\,69$ a.u.}, is deduced from the relativistic recoil for helium atom \cite{r0}, while at $R\rightarrow\infty$ it behaves like $\sim 1/R^4$.

\begin{table}[!htb]
  \caption{The mass-independent nonadiabatic relativistic correction $2\mu_\mathrm{n}\mathcal{E}^{(4,1)}$ (in a.u.)
           for different values of the internuclear distance $R$ (in a.u., for 512 basis size).
           For most of the points, the last digit is uncertain.}
\label{Rs}
\begin{ruledtabular}
\begin{tabular}{lw{9.13}lw{9.13}}
$R$   &\centt{$2\mu_\mathrm{n}\mathcal{E}^{(4,1)}$}& $R$   &\centt{$2\mu_\mathrm{n}\mathcal{E}^{(4,1)}$}\\
  \hline
 0.0  &  -1.079\,69 &    2.1  & -0.122\,89 \\
 0.05 &  -0.761     &    2.15 & -0.107\,12 \\
 0.1  &  -0.511\,6  &    2.2  & -0.091\,78 \\
 0.15 &  -0.384\,7  &    2.3  & -0.062\,21 \\
 0.2  &  -0.383\,85 &    2.4  & -0.034\,67 \\
 0.25 &  -0.433\,56 &    2.5  & -0.008\,80 \\
 0.3  &  -0.505\,79 &    2.6  &  0.014\,69 \\
 0.4  &  -0.630\,36 &    2.7  &  0.035\,88 \\
 0.5  &  -0.702\,97 &    2.8  &  0.054\,42 \\
 0.6  &  -0.725\,41 &    2.9  &  0.069\,94 \\
 0.7  &  -0.713\,63 &    3.0  &  0.082\,16 \\
 0.8  &  -0.681\,06 &    3.2  &  0.097\,14 \\
 0.9  &  -0.637\,11 &    3.4  &  0.099\,47 \\
 1.0  &  -0.587\,82 &    3.6  &  0.091\,98 \\
 1.1  &  -0.537\,03 &    3.8  &  0.078\,60 \\
 1.2  &  -0.486\,52 &    4.0  &  0.063\,08 \\
 1.3  &  -0.437\,70 &    4.2  &  0.048\,65 \\
 1.4  &  -0.390\,85 &    4.4  &  0.036\,55 \\
 1.45 &  -0.368\,28 &    4.6  &  0.027\,37 \\
 1.5  &  -0.346\,24 &    4.8  &  0.020\,32 \\
 1.6  &  -0.304\,09 &    5.0  &  0.015\,45 \\
 1.65 &  -0.283\,70 &    5.2  &  0.011\,82 \\
 1.7  &  -0.263\,96 &    5.4  &  0.009\,19 \\
 1.75 &  -0.244\,62 &    5.6  &  0.007\,30 \\
 1.8  &  -0.225\,83 &    5.8  &  0.005\,90 \\
 1.85 &  -0.207\,76 &    6.0  &  0.004\,91 \\
 1.9  &  -0.189\,95 &    6.5  &  0.003\,13 \\
 1.95 &  -0.172\,52 &    7.0  &  0.002\,05 \\
 2.0  &  -0.155\,56 &    7.5  &  0.001\,43 \\
 2.05 &  -0.139\,04 &    8.0  &  0.001\,11 
\end{tabular}
\end{ruledtabular}
\end{table}

\begin{figure}[!htb]
\includegraphics[scale=0.17]{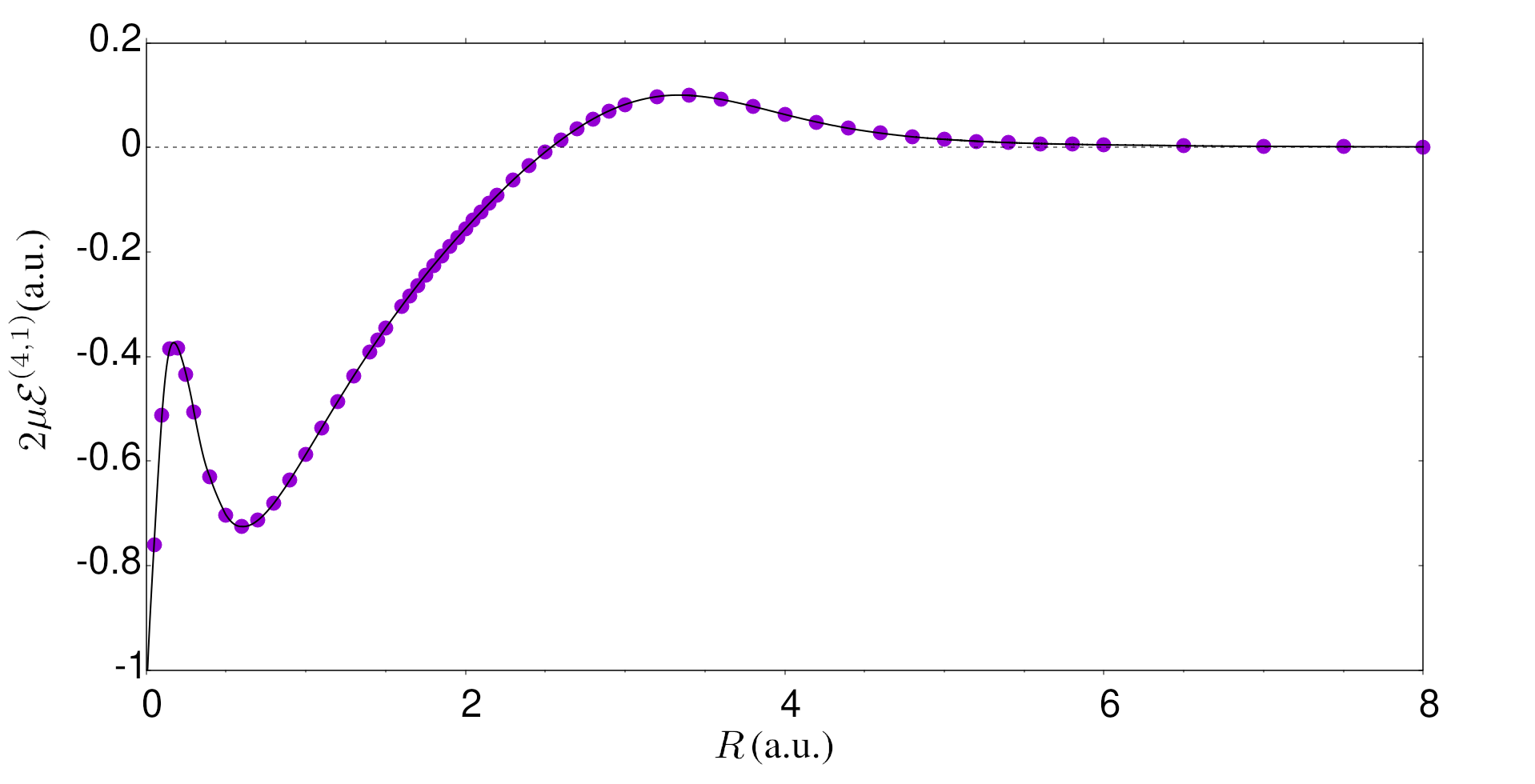}
\caption{Mass-independent nonadiabatic relativistic correction $2\mu_\mathrm{n}\mathcal{E}^{(4,1)}$ for the ground electronic state
         as a function of the internuclear distance $R$.}
\label{FIG}
\end{figure}

\section{Nuclear Schr\"odinger equation}
To obtain the total energy levels one represents $\chi(\vec R)$ as
\begin{align}
  \chi(\vec R) = \frac{\chi(R)}{R}\,Y_{lm}(\vec n)
\end{align}
and solves the radial nuclear Schr\"odinger equation for $\chi(R)$ in the following form
\begin{align}
H_{\rm N}\chi(R) &= E^{(2,0)} \chi(R)\label{nucschro},\\
H_{\rm N}&=-\frac{1}{2\mu_\mathrm{n}}\frac{d^2}{d R^2} + \mathcal{E}(R) +\frac{J(J+1)}{2\mu_\mathrm{n} R^2}, \label{eq59}
\end{align}
where $J$ is the rotational quantum number.
We solve it numerically with a discrete variable representation (DVR) method \cite{dvr}
and obtain a numerical representation of $\chi(R)$ for a specific molecular state.
Note that in some of our previous works we used an adiabatically corrected nuclear function,
which may lead to slight differences between the values presented in this work
and the previous ones. The results of this work clearly demonstrate
that, due to cancellation between different nuclear mass corrections in Eq.\,(\ref{eq62}),
the proper choice for $\chi$ is the BO potential without the adiabatic correction -- Eq. (\ref{eq59}).

The nuclear wave function $\chi(R)$ is subsequently used to calculate the $\alpha^4\,m$ relativistic correction,
according to the following formulas
\begin{align}\label{fullcorr}
E^{(4)}&= E^{(4,0)}+E^{(4,1)},\\
E^{(4,0)}&=\braket{\chi|\mathcal{E}^{(4,0)}(R)|\chi},\\
E^{(4,1)}&=\braket{\chi|\mathcal{E}^{(4,1)}(R)|\chi}+ 2\braket{\chi|\delta\chi}, \label{eq62}
\end{align}
where
\begin{align}
\ket{\delta\chi}&=\mathcal{E}^{(4,0)}(R)\frac{1}{(E^{(2,0)}-H_{\rm N})'}\mathcal{E}^{(2,1)}(R)\ket{\chi}.
\end{align}
The electronic potentials $\mathcal{E}^{(2,1)}(R)$ from Ref. \cite{adiab}, $\mathcal{E}^{(4,0)}(R)$
from Ref. \cite{Mariuszrel}, and $\mathcal{E}^{(4,1)}(R)$ from this work, 
were evaluated on evenly-spaced ($0.05$ a.u.) grid of 200 points and subsequently used in DVR
calculation of matrix elements.
After testing different interpolation schemes, we settled on using the ninth-order piecewise Hermite interpolation.
We observed that the interpolation introduces a relatively significant error to our results,
which could be removed in future via proper analytic fits to $\mathcal{E}^{(4,0)}(R)$ and $\mathcal{E}^{(4,1)}(R)$.

We extrapolate $E^{(4,0)}$, $2\braket{\chi|\delta\chi}$,  and $\braket{\chi|\mathcal{E}^{(4,1)}(R)|\chi}$ separately,
from the results with progressing basis size, and utilize the following model
\begin{align}
E(N)&=\frac{A}{N^k}+E(\infty),
\end{align}
where $N$ is the basis set size and $A$ and $E(\infty)$ are fitted parameters.
We used $k=2$ for $\braket{\chi|\mathcal{E}^{(4,1)}(R)|\chi}$ and $k=3$ in the two other cases.
The choice of $k$ is based on the observation of convergence of individual terms.
The extrapolation error is estimated conservatively to be $50\%$ of the difference between the results
with the two largest basis sets.

\section{Results and discussion}
The total relativistic contribution $E^{(4)}$ to the dissociation energy of H$_2$, HD, and D$_2$
in comparison to fully nonadiabatic naECG calculations from Ref.~\cite{MariuszNa} is shown in Table \ref{TQ}.
The $E^{(4,0)}$ values were obtained as expectation values of the potential from Ref. \cite{Mariuszrel} with a BO nuclear function $\chi$.
We used the recommended CODATA values \cite{codata} for the mass ratios ${m_\mathrm{p}/m_\mathrm{e}=1836.152\,673\,89(17)}$
and ${m_\mathrm{d}/m_\mathrm{e}=3670.482\,967\,85(13)}$,
as well as for the fine-structure ${\alpha=7.297\,352\,566\,4(17)\times10^{-3}}$
and Rydberg ${R_\infty=10\,973\,731.568\,508(65)}$\,m$^{-1}$ constants.
The uncertainty of theoretical results contain the interpolation and extrapolation errors,
as well as the neglected higher-order nonadiabatic corrections estimated by $E^{(4,1)}/\mu_{\rm n}$.
\begin{table}[!htb]
\caption{Convergence of $\braket{\chi|\mathcal{E}^{(4,1)}(R)|\chi}$ contribution to the dissociation energy (in cm$^{-1}$). The remaining components of $E^{(4)}$, Eq.~(\protect\ref{fullcorr}), are also shown for completness.
The uncertainties of the final $E^{(4)}$ values contain an estimate of the higher-order nonadiabatic effects of the order $E^{(4,1)}/\mu_{\rm n}$.}
\label{TQ}
\begin{ruledtabular}
\begin{tabular}{c@{\extracolsep{\fill}}w{3.12}w{2.12}w{2.12}}
\centt{Basis} & \centt{H$_2$} & \centt{HD} & \centt{D$_2$} \\
\hline\\[-1.5ex]
128                                 &  0.002\,376\,07 &   0.001\,794\,00 & 0.001\,205\,590\\
256                                 &  0.002\,370\,19 &   0.001\,789\,61 & 0.001\,202\,694\\
512                                 &  0.002\,368\,67 &   0.001\,788\,48 & 0.001\,201\,951\\
$\infty$                            &  0.002\,368(1)      &   0.001\,788(1)      & 0.001\,201\,7(4)\\
$2\braket{\chi|\delta\chi}$         & -0.000\,451\,1  &  -0.000\,342\,0  &-0.000\,230\,9\\
\hline$E^{(4,1)}$                   &  0.001\,917(1)      &   0.001\,446(1)      & 0.000\,970\,8(4) \\
$E^{(4,0)}$                         & -0.533\,130(1)      &  -0.531\,334(1)      &-0.529\,179(1) \\
\hline$E^{(4)}$                     & -0.531\,213(2)      &  -0.529\,888(2)      &-0.528\,208(1)\\
naECG \cite{MariuszNa}              & -0.531\,215\,6(5)   &  -0.529\,887\,5(2)   &-0.528\,206\,1(1)\\
Difference                          &  0.000\,003(2)      &  -0.000\,001(2)      &-0.000\,002(1)
\end{tabular}
\end{ruledtabular}
\end{table}

A good agreement between results of the naECG from Ref. \cite{MariuszNa} and of NAPT obtained here
for the ground molecular state (see Table II) justifies the perturbative approach, the main advantage of which is
the common potential ${\cal E}^{(4,1)}(R)$ for all the rovibrational states of all isotopes of molecular hydrogen
in the ground electronic state.

The $\alpha^n\,m$ contributions with $n=5,6,7$ in the following tables
are expectation values of the potentials from the references given
in the last column, with the BO nuclear wave function $\chi$.
While all the $\alpha^n\,m$ contributions are calculated according to known formulas,
the formula for $\alpha^7\,m$ correction is yet unknown.
Their values presented in Tables~\ref{TD}-\ref{D2tr} are only estimates, hence the $50\%$ error, based on the leading term analogous as in atomic hydrogen, namely \cite{eides:01}
\begin{align}
  H^{(7)}\approx&-\left[\delta^3(r_{1A})+\delta^3(r_{2A})+\delta^3(r_{1B})+\delta^3(r_{2B})\right]
  \nonumber \\ &\times  \mathrm{ln}^2(\alpha^{-2}). 
\end{align}
The expectation values of the Dirac $\delta$ were taken from Ref. \cite{Mariuszrel}.
They were used also in the evaluation of the correction due to the finite nuclear size \cite{eides:01}
\begin{align}
  E_\mathrm{FS}=&\alpha^4\,
  \frac{2\pi}{3}\left[\delta^3(r_{1A})+\delta^3(r_{2A})+\delta^3(r_{1B})+\delta^3(r_{2B})\right]
  \nonumber \\ &\times \frac{(r_A^2+r_B^2)}{2}\,,\label{FS}
\end{align}
where $r_{A/B}$ is the root mean square charge radius of the $A/B$ nucleus. The higher-order
effects due to the nuclear size or nuclear polarizability are negligible at the current
precision level, which we know from the atomic hydrogen and deuterium \cite{eides:01}.

In Table \ref{TD}, we present the theoretical predictions for the dissociation energy of $v=0$, $J=1$ state of  H$_2$,
and two selected transitions in comparison to the most accurate experimental data. 
We find an agreement for the
dissociation energy of the former level and for the $(3,5)\rightarrow(0,3)$ transition energy,
whereas for the $(1,0)\rightarrow(0,0)$ transition a $2\,\sigma$ disagreement persists.
For this reason, the experimental value for this transition should be verified.
\begin{table*}[!htb]
\renewcommand{\arraystretch}{1.3}
\caption{Contributions to the dissociation energy of
  the first rotationally excited level $(v,J) = (0,1)$ and 
  to two selected transitions in H$_2$ (in cm$^{-1}$), in comparison to experimental values.
  ${E}_\mathrm{rel}^{(2)}\sim \alpha^6\,m$ is a second order correction due to relativistic BO potential,
  which in former works was automatically included in $\alpha^4\,m$;
  $E_\mathrm{FS}$ is the finite nuclear size correction with $r_p = 0.84087(39)$ fm \cite{Antognini:13}.}
\label{TD}
\begin{ruledtabular}
\begin{tabular}{lw{6.11}w{6.11}w{6.11}l}
\centt{Contribution/$(v,J)$}& \centt{$(0,1)$} & \centt{$(3,5)\rightarrow(0,3)$}& \centt{$(1,0)\rightarrow(0,0)$} & References \\
\hline
$\alpha^2\,m$  & 36\,000.312\,485\,66(6)&12\,559.749\,918\,95(8)  & 4\,161.163\,977\,09(6)  & \cite{komasa_2018}, \cite{Mariuszrel}, \cite{Mariuszrel} \\
$\alpha^4\,m$  &      -0.533\,796(2)    &      0.065\,878(8)     &        0.023\,554(2)     & \cite{adiab} + \cite{Mariuszrel} + this work \\
$\alpha^5\,m$  &      -0.194\,00(21)    &     -0.065\,81(7)       &      -0.021\,32(2)      & \cite{Mariuszrel} + \cite{Grzegorz2009}\\
$\alpha^6\,m$  &      -0.002\,058(6)    &     -0.000\,599(1)      &      -0.000\,192        & \cite{ma6} \\
${E}_\mathrm{rel}^{(2)}$ & 0.000\,009& 0.000\,004&0.000\,001 & \cite{Mariuszrel}\\
$\alpha^7\,m$  &       0.000\,117(59)   &      0.000\,037(19)     &       0.000\,012(6)     & \cite{Mariuszrel}\\
$E_\mathrm{FS}$  &      -0.000\,031       &     -0.000\,010         &      -0.000\,003        & \cite{Mariuszrel}\\
\hline Total          & 35\,999.582\,73(22)    &12\,559.749\,42(7)       &  4\,161.166\,03(2)      & \\
Experiment     & 35\,999.582\,894(25)   &12\,559.749\,52(5)       &  4\,161.166\,36(15)     & \cite{Ubachs2}, \cite{exp2}, \cite{Niu}\\
Difference     &      -0.000\,16(22)    &     -0.000\,10(9)       &      -0.000\,33(15)     & \\  
\end{tabular}
\end{ruledtabular}
\end{table*}

Because all the relativistic and QED corrections are calculated through effective potentials,
they can be employed to obtain all the rovibrational  energies of all isotopes of the hydrogen molecule
for the ground electronic state. In Tables \ref{HDtr} and \ref{D2tr} we present results 
for a selection of transitions in HD and D$_2$ which have been measured with a high accuracy.
In general, we observe very good agreement between theoretical and experimental data,
except for the series of $R_2(J)$ transitions in HD, presented in the lower panel of Table IV,
which requires further investigations.

\begin{table*}[!htb]
\renewcommand{\arraystretch}{1.3}
\caption{Contributions to selected transitions in HD (in cm$^{-1}$).
  ${E}_\mathrm{rel}^{(2)} \sim \alpha^6\,m$ is a second order correction due to relativistic BO potential,
  which in former works was automatically included in $\alpha^4\,m$.
  $E_\mathrm{FS}$ is the finite nuclear size correction with $r_p = 0.84087(39)$ fm \cite{Antognini:13}
  and $r_d =2.12771(22)$ fm \cite{Pohl:16}. There are two additional measurements of  $(2,2)\rightarrow(0,1)$ transition,
  namely $7\,241.849\,386(3)$ cm$^{-1}$ \cite{Tao} and  $7\,241.849\,345\,6(32)$ \cite{Gianfrani}
  which are in disagreement with that in the table.}
\label{HDtr}
\begin{ruledtabular}
\begin{tabular}{lw{6.11}w{6.11}w{6.11}l}
\centt{Contribution/$(v,J)$}&\centt{$(1,0)\rightarrow(0,0)$} & \centt{$(0,1)\rightarrow(0,0)$} & \centt{$(1,1)\rightarrow(0,1)$} & References \\
\hline
$\alpha^2\,m$  &3\,632.158\,204\,27(1) &89.226\,757\,95(1) & 3\,628.302\,279\,75(1)               & \cite{pccp2018}\\
$\alpha^4\,m$  &     0.020\,999(1)& 0.001\,950\,56(1)&0.020\,856(1)                 & \cite{adiab} + \cite{Mariuszrel} + this work\\
$\alpha^5\,m$  &    -0.018\,64(2)      &-0.000\,770\,9(6)     &     -0.018\,60(2)                    & \cite{Mariuszrel} + \cite{Grzegorz2009}\\
$\alpha^6\,m$  &    -0.000\,168        &-0.000\,006\,74(1)        &     -0.000\,168                      & \cite{ma6}\\
${E}_\mathrm{rel}^{(2)}$ & 0.000\,001 & 0.000\,000\,06& 0.000\,001   & \cite{Mariuszrel}\\
$\alpha^7\,m$  &     0.000\,010(5)      & 0.000\,000\,43(22)    &      0.000\,010(5)                      & \cite{Mariuszrel}\\
$E_\mathrm{FS}$  &    -0.000\,010        &-0.000\,000\,43    &     -0.000\,010                      & \cite{Mariuszrel}\\
\hline Total          &3\,632.160\,40(2)      & 89.227\,930\,9(6)    & 3\,628.304\,37(2)                   &\\
Experiment            &3\,632.160\,52(22)     & 89.227\,931\,6(8)    & 3\,628.304\,50(22)              &\cite{Niu}, \cite{Drouin}, \cite{Niu}\\
Difference     &    -0.000\,12(22)     & -0.000\,000\,7(10)  &     -0.000\,13(22)         &\\
\hline\hline\centt{Contribution/$(v,J)$}&\centt{$(2,2)\rightarrow(0,1)$} & \centt{$(2,3)\rightarrow(0,2)$} & \centt{$(2,4)\rightarrow(0,3)$} & References \\
\hline
$\alpha^2\,m$  &                                 7\,241.846\,168\,22(2)  & 7\,306.479\,554\,52(2)   & 7\,361.899\,285\,85(1)  & \cite{pccp2018}\\
$\alpha^4\,m$  &                                      0.040\,927(4) &      0.041\,927(3)  &      0.042\,559(3) & \cite{adiab} + \cite{Mariuszrel} + this work \\
$\alpha^5\,m$  &                                     -0.037\,46(3)       &     -0.037\,97(3)        &     -0.038\,38(3)       & \cite{Mariuszrel} + \cite{Grzegorz2009}\\
$\alpha^6\,m$  &                                     -0.000\,339         &     -0.000\,343          &     -0.000\,347         & \cite{ma6}\\
${E}_\mathrm{rel}^{(2)}$ &      0.000\,002      &      0.000\,002       &      0.000\,002      & \cite{Mariuszrel}\\
$\alpha^7\,m$  &                                      0.000\,021(11)     &      0.000\,021(11)      &      0.000\,022(11)     & \cite{Mariuszrel}\\
$E_\mathrm{FS}$  &                                     -0.000\,021         &     -0.000\,021          &     -0.000\,021         & \cite{Mariuszrel}\\
\hline Total   &                                 7\,241.849\,30(3)       & 7\,306.483\,17(3)        & 7\,361.903\,12(3)       & \\
Experiment   &                                  7\,241.849\,356\,16(67) & 7\,306.483\,227\,84(93)  & 7\,361.903\,178\,73(93) & \cite{Ubachs1}\\
Difference     &                            -0.000\,06(3)       &     -0.000\,06(3)        &     -0.000\,06(3)       & \\
\end{tabular}
\end{ruledtabular}
\end{table*}

\begin{table*}[!htb]
\renewcommand{\arraystretch}{1.3}
\caption{Contributions to selected transitions in D$_2$ (in cm$^{-1}$). ${E}_\mathrm{rel}^{(2)} \sim \alpha^6\,m$
  is a second order correction due to relativistic BO potential, which in former works
  was automatically included in $\alpha^4\,m$.
  $E_\mathrm{FS}$ is the finite nuclear size correction with $r_d =2.12771(22)$ fm \cite{Pohl:16}.}
\label{D2tr}
\begin{ruledtabular}
\begin{tabular}{lw{6.11}w{6.11}w{6.11}l}
\centt{Contribution/$(v,J)$} & \centt{$(2,4)\rightarrow(0,2)$}& \centt{$(1,0)\rightarrow(0,0)$} & \centt{$(1,2)\rightarrow(0,2)$} & References \\
\hline
$\alpha^2\,m$        & 6\,241.120\,920(1)       &  2\,993.614\,856\,52(4)   & 2\,987.291\,387\,6(2)  & \cite{Wcislo}, this work, this work\\
$\alpha^4\,m$        &      0.040\,173\,9(15)   &       0.017\,732\,2(2)    &      0.017\,498& \cite{adiab} + \cite{Mariuszrel} + this work \\
$\alpha^5\,m$        &     -0.033\,167(18)      &      -0.015\,397(8)       &     -0.015\,33(1)    & \cite{Mariuszrel} + \cite{Grzegorz2009}\\  
$\alpha^6\,m$        &     -0.000\,298\,9(2)      &      -0.000\,138\,7(1)    &     -0.000\,138    & \cite{ma6}\\ 
${E}_\mathrm{rel}^{(2)}$ & 0.000\,001\,9& 0.000\,000\,9   &      0.000\,001& \cite{Mariuszrel}\\
$\alpha^7\,m$        &      0.000\,019(10)      &       0.000\,008\,6(43)   &      0.000\,009(4)&\cite{Mariuszrel} \\
$E_\mathrm{FS}$        &     -0.000\,031\,5       &      -0.000\,014\,6       &     -0.000\,015    &\cite{Mariuszrel}\\
\hline Total         & 6\,241.127\,617(21)      &  2\,993.617\,048(9)      & 2\,987.293\,41(1)      &\\
Experiment           & 6\,241.127\,647(11)      &  2\,993.617\,06(15)       & 2\,987.293\,52(15) &\cite{Wcislo}, \cite{Niu}, \cite{Niu}\\
Difference           &     -0.000\,030(24)      &      -0.000\,01(15)       &     -0.000\,11(15)
\end{tabular}
\end{ruledtabular}
\end{table*}

\section{Conclusions}
The main achievement of this work is a significant reduction of the contribution to the total
error budget coming from the nonadiabatic relativistic (recoil) effects.
As a result, the current main source of theoretical uncertainty
is the unknown combined QED and nonadiabatic correction,
which is estimated by the ratio of the electron to the reduced nuclear masses $1/\mu_{\rm n}$.
We have already undertaken calculation of this missing term.
Once this contribution is known, the main uncertainty will come from the approximate value of the $\alpha^7\,m$ term, accurate calculation 
of which is very challenging. If these calculations are accomplished,
together with the leading nonadiabatic $\alpha^6\,m$ correction,
one can use precisely measured transitions in the molecular hydrogen to determine fundamental physical constants,
such as the proton-electron mass ratio or the nuclear charge radii, for which discrepant values
have been obtained in the literature.

\section*{Acknowledgements}
This  project is supported by a National  Science  Centre, Poland grants No. 2017/25/N/ST4/00594 (P.C. and K.P),
2016/23/B/ST4/01821 (M.P.) and 2017/25/B/ST4/01024 (J.K.). The computations were performed in \'{S}wierk
Computing Centre and Pozna\'{n} Supercomputing and Networking Center.
We would like to express our gratitude to Grzegorz \L ach for inspiring discussions and to Magdalena Zientkiewicz
-- for her help with the choice of the model for the extrapolation of our results.


\begin{thebibliography}{99}
\bibitem{NRQED}W. E. Caswell, G. P. Lepage, Phys. Lett. B {\bf{167}}, 437 (1986). 
\bibitem{NAPT}K. Pachucki, J. Komasa, J. Chem. Phys. {\bf{143}}, 034111 (2015).
\bibitem{komasa_2018} K. Pachucki and J. Komasa, Phys. Chem. Chem. Phys. {\bf 20}, 247 (2018).
\bibitem{Yan}L. M. Wang and Z.-C. Yan, Phys. Rev. A {\bf 97}, 060501(R) (2018).
\bibitem{MariuszNa} M. Puchalski, A. Spyszkiewicz, J. Komasa, and K. Pachucki, Phys. Rev. Lett. {\bf{121}}, 073001 (2018).
\bibitem{Yan2}L. M. Wang and Z.-C. Yan, Phys. Chem. Chem. Phys. {\bf 20}, 23948 (2018).
\bibitem{adiab}K. Pachucki, J. Komasa, J. Chem. Phys. {\bf{141}}, 224103 (2014). 
\bibitem{Mariuszrel}M. Puchalski, J. Komasa, K. Pachucki,  Phys. Rev. A {\bf{95}}, 052506 (2017).
\bibitem{Drachman} R. J. Drachman, J. Phys. B {\bf{14}}, 2733 (1981), 
\bibitem{Cencek} K. Pachucki, W. Cencek and J, Komasa, J. Chem. Phys. {\bf 122}, 184101 (2005).
\bibitem{gaussext} K. Pachucki, M. Puchalski and V.~A. Yerokhin, Comp. Phys. Comm. {\bf 185}, 2913  (2014).
\bibitem{ma6}M. Puchalski, J. Komasa, P. Czachorowski, K. Pachucki,  Phys. Rev. Lett. {\bf{117}}, 263002 (2016).
\bibitem{r0} K. Pachucki, V. Patk\'{o}\v{s}, V. A. Yerokhin, Phys. Rev. A {\bf{95}}, 062510 (2017).
\bibitem{dvr}D. T. Colbert and W. H. Miller, J. Chern. Phys. 96, 1982 (1992).
\bibitem{codata} P. J. Mohr, D. B. Newell, and B. N. Taylor, Rev. Mod. Phys. {\bf{88}}, 035009 (2016).
\bibitem{eides:01} M.~I. Eides, H.~Grotch, and V.~A. Shelyuto, \newblock Phys. Rep. {\bf 342}, 63 (2001).
\bibitem{Antognini:13} A. Antognini, F. Nez, K. Schuhmann, F. D. Amaro, F. Biraben, J. M. R. Cardoso, D. S. Covita, A. Dax, S. Dhawan, M. Diepold, {\em et al.}, Science {\bf{339}}, 417 (2013).
\bibitem{Grzegorz2009} K. Piszczatowski, G. \L ach, M. Przybytek, J. Komasa, K. Pachucki and B. Jeziorski,
                 J. Chem. Theory Comput., {\bf 5}, 3039 (2009).
\bibitem{Ubachs2}C. Cheng, J. Hussels, M. Niu, H.L. Bethlem, K.S.E. Eikema, E.J. Salumbides, W. Ubachs, M. Beyer, N. Hoelsch,
                 J. A. Agner, F. Merkt, L.G. Tao, S.M. Hu, C. Jungen, Phys. Rev. Lett. {\bf{121}}, 013001 (2018).
\bibitem{exp2} C.-F. Cheng, Y. R. Sun, H. Pan, J. Wang, A.-W. Liu, A. Campargue, and S.-M. Hu, Phys. Rev. A {\bf{85}}, 024501 (2012).
\bibitem{Niu}M. Niu, E.J. Salumbides, G.D. Dickenson, K.S.E. Eikema, W. Ubachs, J. Mol. Spectr. {\bf{300}}, 44-54 (2014).
\bibitem{Pohl:16} R. Pohl {\em et al.}, Science {\bf 353}, 669 (2016).
\bibitem{Tao}L.-G. Tao, A.-W. Liu, K. Pachucki, J. Komasa, Y. R. Sun, J. Wang, S.-M. Hu, Phys. Rev. Lett. {\bf{120}}, 153001 (2018).
\bibitem{Gianfrani} E. Fasci, A. Castrillo, H. Dinesan, S. Gravina, L. Moretti, and L. Gianfrani, Phys. Rev. A {\bf{98}}, 022516 (2018).
\bibitem{pccp2018} K. Pachucki and J. Komasa, Phys. Chem. Chem. Phys. {\bf 20}, 26297 (2018). 
\bibitem{Drouin}B. J. Drouin, S. Yu, J. C. Pearson, H. Gupta, J. Mol. Struct. {\bf{1006}}, 2-12 (2011).
\bibitem{Ubachs1}F. M. J. Cozijn, P. Dupr\'{e}, E. J. Salumbides, K. S. E. Eikema, and W. Ubachs, Phys. Rev. Lett. {\bf{120}}, 153002 (2018).
\bibitem{Wcislo} P. Wcis\l{}o, F. Thibault, M. Zaborowski, S. W\'{o}jtewicz, A. Cygan, G. Kowzan, P. Mas\l{}owski,
                 J. Komasa, M. Puchalski, K. Pachucki, R. Ciury\l{}o, D. Lisak, J. Quant. Spectrosc. Radiat. Transf. 213, {\bf{41}} (2018).
\end{thebibliography}
\end{document}